\documentclass[12pt]{article}
\usepackage{graphicx}
\usepackage{epstopdf}
\usepackage{amsfonts}
\usepackage{amssymb}
\usepackage[footnotesize]{caption2}
\usepackage{mathrsfs}
\usepackage{amsmath}
\usepackage[numbers, sort&compress]{natbib}

\usepackage{authblk}

\begin{document}
\renewcommand{\thefootnote}{\fnsymbol{footnote}}
\title{One-dimension quantum systems in the framework of the most general deformation GUP form}
\author{\small Ying-Jie Zhao\thanks{E-mail address: xiangyabaozhang@qq.com} ,
\small Yan-Fang Ji\thanks{E-mail address: jiyanfang88@qq.com} ,
\small Guang-Rui Yao\thanks{E-mail address: coldstar\_ygr@qq.com} ,
\small Xiaojie Li\thanks{E-mail address: 13256199291@163.com} ,
\small Muddasir Hanif\thanks{E-mail address: muddasirhanif@yahoo.com}}

\affil{\small Qilu Institute of Technology, 3028 Jingshi East Road, Jinan City, Shandong Province, China}

\renewcommand{\Authands}{, }
\date{}
\maketitle
\begin{abstract}

\setlength{\parindent}{0pt} \setlength{\parskip}{1.5ex plus 0.5ex minus 0.2ex} %\noindent

This paper provides a concise overview of the comprehensive version of the Generalized Uncertainty Principle (GUP) derived from nonlocal quantum mechanics and extensively addressed by S. Masood, {\em et al}. We  utilize the specific constraint of this GUP formulation to compute the energy correction and the modified wave function for the linear potential in one dimension. In addition, we  examine the impact of deformation on a one-dimensional delta potential well and a one-dimensional delta potential barrier. Furthermore, we specifically focus on the impact of deformation on hydrogen atoms existing in one dimension, and we conduct a distinct investigation into the Stark effect on the atom.
\vskip 10pt
\noindent


\end{abstract}

\thispagestyle{empty}

\newpage
\setcounter{page}{1}

\section{Introduction}
Many approaches to unify the quantum mechanics and general relativity, including string theory \cite{s1, s2, s3, s3_1, s3_2, s4}, loop quantum gravity \cite{s5}, and quantum geometry \cite{s6} have attracted much attention in recent years. Almost all these proposals, together with some experiments \cite{s7}, support a minimal length of the order of Planck scale and a modification of the Heisenberg uncertainty principle (HUP) to the so-called generalized uncertainty principle (GUP)\cite{s8, s9, s10, s11, s12}.
In the framework of GUP, contrary to the HUP's one, a lower bound of the measurable length of the order of the Planck scale $10^{-35}m$ naturally appears in the spacetime\cite{s10, s11, s12, s13}.

Generalized uncertainty principle, one of the approaches in adding quantum effects on gravity systems, have received much attention and several achievements have been made \cite{s15, s15_1, s16, s16_1, s16_2, s16_3, s16_4, s17, s17_1, s17_2, s17_3, s17_4, s19, s19_1, s19_2, s19_3, s19_4, s19_5, s19_6, s19_7, s19_8, s19_9, s19_10, s19_11}. For instance, in the framework of GUP, the existence of a minimal length prohibits the wholly radiation of a black hole and there finally remains a remnant that the information loss paradox may be avoided.

Various typical forms of GUP deforming the position-momentum commutation relations  have been mentioned by researchers. One is the quadratic form GUP \cite{s8} proposed by KMM in which the commutators of position and momentum operators contain a additional quadratic term in momentum, deforms the momentum to
\begin{eqnarray}
p_i \rightarrow  \tilde{p}_i = p_i( 1+ \lambda p^j p_j ),
\end{eqnarray}
Another is the linear form GUP proposed by S. Das \cite{s16, s16_1,s16_2,s16_3,s16_4} that deforms the momentum to
\begin{eqnarray}
p_i \rightarrow  \tilde{p}_i = p_i \left( 1+ \lambda_1 \sqrt{p^j p_j}+2 \lambda_2 p^j p_j \right).
\end{eqnarray}
Besides, in our previous work \cite{s20, yys}, we have introduced an improved exponential form GUP and obtain the cosmological constant that coincides exactly with the experimental value provided by the Planck 2013 results \cite{Planck} by choosing an appropriate index $n$ in our GUP while considering the UV/IR mixing effect. Moreover, we have investigated the maximally localized states, the corresponding quasi-position wave functions, and the scalar product
of these wave functions, derived the corrected thermodynamic quantities of the Schwardzschild black hole with or without considering the UV/IR mixing effect. We have analyzed these results in different cases of index $n$ and made some interesting conclusions.

However, all these forms of the GUP in previous papers are only particular. In fact, it is necessary to find out the most general deformation of the Heisenberg algebra containing the inverse powers of the momentum operators motivated by GUP. The non-locality method \cite{snonlocalty, snonlocalty_1, snonlocalty_2, snonlocalty_3, snonlocalty_4}, as one of the approaches to obtain the deformation of the momentum, has been widely utilized to achieve the goal. In the method, the deformation of the momentum operator can produce linear terms of momentum operator giving rise to a discretization of space. In the framework of the GUP deformation, the authors of the previous paper \cite{smostgeneralGUP} applied the deformation on a harmonic oscillator and observed that there was no correction to the energy in the first order but the second order. They also analyzed the Landau levels and found that the corrections of Landau levels existed at first order. Besides, the wave functions of the Lamb shift altered at first order of the perturbation theory, and the tunneling current of a barrier potential had changed.

The paper is organized as follows. In Section $2$ we briefly review the most general form of GUP motivated from nonlocal quantum mechanics discussed in detail in refs.\cite{smostgeneralGUP}. In section $3$, we apply the special limit of this GUP form to the one-dimension linear potential to calculate the energy corrections and the modified wave function. In section $4$, we discuss one dimensional delta potential well. In section $5$, we then focus on one dimensional delta potential barrier. In section $6$ and $7$ we also pay attention to one dimensional hydrogen atom and investigate the Stark effect on the atom separately. At the end, we make a short conclusion in section $8$.
\\
\\
\\
\section{Non-locality}
The non-locality quantum mechanics \cite{snonlocalty, snonlocalty_1, snonlocalty_2, snonlocalty_3, snonlocalty_4} can generate the most general deformation of the momentum operator including inverse powers terms.
For instance, as an application to quantum mechanics, the non-local deformation of Schr\"{o}edinger equation can be written as
\begin{eqnarray}
i \hbar {\partial_t} \psi(x) +\frac{\hbar^2}{2m} {\partial^i}{\partial_i}\psi(x) -V(x) \psi(x) = \int d^3 y K(x,  y) \psi(y),
\end{eqnarray}
here $K(x,  y)$ denotes the non-local operator. Moreover, non-local deformation of field theory \cite{smostGUPField, smostGUPField_1, smostGUPField_2, smostGUPField_3} and  non-local deformation of gravity \cite{smostGUPGravity, smostGUPGravity_1, smostGUPGravity_2, smostGUPGravity_3} has also been discussed and lots of results have been gained. In the refs.\cite{smostgeneralGUP}, S.Masood, {\em et.al.}  applies the non-locality  method to a simple massless scalar field theory,
\begin{eqnarray}
\hbar^2 \partial^\mu \partial_\mu \psi(x) = 0.
\end{eqnarray}
By adding a non-local source term on the RHS of the equation ($\lambda$ is the coupling parameter measuring the coupling of the non-local part of the theory)
\begin{eqnarray}
\hbar^2 \partial^\mu \partial_\mu \psi(x) = \lambda \int d^4 y G(x,  y) \psi(y),
\end{eqnarray}
where
\begin{eqnarray}
G(x-y)  = \int\frac{d^4 p}{(2\pi)^4}\frac{1}{p^2} e^{i p(x-y)},
\end{eqnarray}
and  get the equation is
\begin{eqnarray}
& &- \int d^4 y \left[\delta(x-y) \hbar^2  \partial^\mu \partial_\mu \psi(y) - \lambda G(x-y) ] \psi(y) \right]    \nonumber \\
&=&  \int\frac{d^4 p d^4 y}{(2\pi)^4}\left[\left(p^2 +\frac{\lambda}{p^2}\right) e^{i p(x-y)}\right]\psi(y)    \nonumber \\
&=& 0.                         \label{nonlocal1}
\end{eqnarray}
Eq.(\ref{nonlocal1})  implies that the usual  scalar field  can be deformed via the momentum transformation
 \begin{eqnarray}
p^2 \rightarrow \tilde{p}^2 =p^2 +\frac{\lambda}{p^2},
\end{eqnarray}
which includes the inverse power term of momentum operators.  Here only  the spatial deformation is considered and hence the most  general form of GUP
can be  constructed when  all positive and negative powers of momentum operators included,
 \begin{eqnarray}
p_i\rightarrow\tilde{p}_i = p_i \left[1\pm\sum_r \lambda_{1r}(p^j p_j)^{r/2}\pm\sum_r \lambda_{2r}(p^j p_j)^{-r/2}\right]. \label{XandP}
\end{eqnarray}

For simplicity, we consider the simple limit in one-dimension momentum space, and  this   momentum  deformation  can be represented as
\begin{eqnarray}
\tilde{p} = p \left(1+\frac{\lambda}{p}\right),            \label{XandP}
\end{eqnarray}
which deforms common Hamiltonian of the system  as
\begin{eqnarray}
H = \frac{p^2}{2m} + V(x) = \frac{p^2}{2m}+ \frac{\lambda p}{m} +V(x).
\end{eqnarray}

Up to now a local deformation of all one dimensional quantum mechanical systems has been generated by nonlocal deformation of the momentum operator. We see that though the momentum operator includes nonlocal terms, the Hamiltonian for one dimensional quantum mechanical systems is conversely local.
\\
\\
\\
\section{One-dimension linear potential}
In this section  we will discuss the effect of this deformation on the one-dimension linear potential.
Now let us consider a one dimensional linear potential, which has the form ($F$ is a positive constant)
 \begin{eqnarray}
 V(x) =\left\{
\begin{aligned}
 F x ,   x>0 \\
 +\infty,   x<0
 \end{aligned}
\right.
\end{eqnarray}
In consequence of the deformation of the momentum operator $p\rightarrow p\left(1+\lambda/p \right)$, the corrected Hamiltonian governing the motion of the  particle of mass $m$ in  a one-dimension linear potential is shown as
\begin{eqnarray}
H = \frac{p^2}{2m} + F x \rightarrow \frac{p^2}{2m} + \frac{\lambda p}{m} + F x.
\end{eqnarray}
and the deformed Schr\"{o}dinger equation reads
\begin{eqnarray}
\frac{\hbar^2}{2 m} \psi''(x) + i\hbar\frac{\lambda}{ m} \psi'(x) + \left(E - F x \right)\psi(x) =0 . \label{linearequation1}
\end{eqnarray}
with $\psi(x)\rightarrow 0 $  when $x \rightarrow 0$ or $x \rightarrow +\infty$.
The above equation  has a solution composed of the first and second kinds of Airy functions ($C_1$ and $C_2$ are constant)£¬
\begin{eqnarray}
\psi(x)= e^{-\frac{i \lambda x}{\hbar}} \left[ C_1  Ai\left(\frac{-2 m E +2 m F x -\lambda^2}{(2mF\hbar)^{2/3}}\right)+ C_2
Bi\left(\frac{-2 m E +2 m F x -\lambda^2}{(2mF\hbar)^{2/3}}\right)\right].
\end{eqnarray}
The boundary condition  $\psi(x)\rightarrow 0$ when  $x\rightarrow \infty$ indicates that $Bi$ should be discarded and the wave function is given by
\begin{eqnarray}
\psi(x)= C_1 e^{-\frac{i \lambda x}{\hbar}}  Ai\left(\frac{-2 m E +2 m F x -\lambda^2}{(2mF\hbar)^{2/3}}\right).
\end{eqnarray}
And further the other boundary condition  $\psi(x)\rightarrow 0$ when  $x\rightarrow 0$ gives rise to an Airy equation,
\begin{eqnarray}
Ai\left(\frac{-2 m E -\lambda^2}{(2mF\hbar)^{2/3}}\right) = 0,
\end{eqnarray}
which implies the energy level  should be quantized,
\begin{eqnarray}
\frac{-2 m E_n -\lambda^2}{(2mF\hbar)^{2/3}} = a_n,
\end{eqnarray}
where $a_1= -2.33810, a_2=-4.08794, a_3=-5.52055, a_4= -6.78670, a_5=-7.94413,\cdots$ is the zeros of the first kind of Airy function.
Using this result the energy can be derived as
\begin{eqnarray}
E_n = - \left(\frac{F^2 \hbar^2}{2m}\right)^{1/3}a_n - \frac{\lambda^2}{2m}.
\end{eqnarray}
At last, without loss of generality we choose the wave function
\begin{eqnarray}
\psi(x) = \frac{(\frac{2 m F}{\hbar^2})^{1/6}}{|Ai'(a_n)|}  e^{-\frac{i \lambda x}{\hbar}}  Ai\left( a_n + \left(\frac{2 m F}{\hbar^2}\right)^{1/3} x \right),
\end{eqnarray}
here $Ai'(x)$ is the derivative of $Ai(x)$.
\\
\\
\\
\section{One-dimension delta potential well}
In this section we will derive the effect of the the deformation on a One-dimension delta potential well.
The one-dimension  delta potential well takes the following form
\begin{eqnarray}
V(x) = - V \delta(x),
\end{eqnarray}
where $\delta(x)$ is the Dirac delta function and V is a positive constant. The deformed Hamiltonian for the delta potential well  takes the form
\begin{eqnarray}
H = \frac{p^2}{2m} + V(x) = \frac{p^2}{2m} + \frac{\lambda p}{m} -V \delta(x)
\end{eqnarray}
and the deformed Schr\"{o}edinger equation is shown as follows
\begin{eqnarray}
\frac{\hbar^2}{2 m} \psi''(x) + i\hbar\frac{\lambda}{ m} \psi'(x) + \left[ V \delta(x) +E \right]\psi(x) =0 \label{deltapotentialwellSch}
\end{eqnarray}
Intrgrating the LHS of the Eq.(\ref{deltapotentialwellSch}) over $x$ from $- \varepsilon$ to $+\varepsilon$ and afterwards let $\varepsilon \rightarrow 0$, with the sifting property of delta function $\delta(x)$
\begin{eqnarray}
\int_{-\varepsilon}^{+\varepsilon}\psi(x) \delta(x) dx = \psi(0),
\end{eqnarray}
from Eq.(\ref{deltapotentialwellSch}) we have the step equation of $\psi'(x)$
\begin{eqnarray}
 \psi'\left(0^+\right)-  \psi'\left(0^-\right) = - \frac{2 m V }{\hbar^2} \psi\left(0\right). \label{connectionequation1}
\end{eqnarray}
This indicate that the wave function $\psi(x)$  continues at the singularity of the potential $V(x)$ while $\psi'(x)$ does not. In the region of $x \neq 0$  the  potential $V(x) =0$ thus the equation reads as
\begin{eqnarray}
\frac{\hbar^2}{2 m} \psi''(x) + i\hbar\frac{\lambda}{ m} \psi'(x) +    E  \psi(x) =0.
\end{eqnarray}
The solution of the above equation is
\begin{eqnarray}
 \psi(x) =
A e^{-\frac{i \lambda x}{\hbar}} e^{-\frac{x}{\hbar}\sqrt{-2 m E - \lambda^2}}+ B e^{-\frac{i \lambda x}{\hbar}} e^{\frac{x}{\hbar}\sqrt{-2 m E - \lambda^2}}.
 \end{eqnarray}
Taking the boundary condition $x \rightarrow  \pm\infty, \psi(x)\rightarrow 0$ into consideration , we obtain
$A=0$ when $x<0$ and $B=0$ when $x>0$. So we get
\begin{eqnarray}
 \psi(x) =\left\{
\begin{aligned}
&& A e^{-\frac{i \lambda x}{\hbar}} e^{-\frac{x}{\hbar}\sqrt{-2 m E - \lambda^2}},   x>0 \\
&& A e^{-\frac{i \lambda x}{\hbar}} e^{\frac{x}{\hbar}\sqrt{-2 m E - \lambda^2}} ,   x<0 \label{deltapotentialwellequation1}
\end{aligned}
\right.
\end{eqnarray}
Applying the Eq.(\ref{connectionequation1}) to the Eq.(\ref{deltapotentialwellequation1}) one can calculate the corrected bound state energy of the delta potential well
\begin{eqnarray}
E = - \frac{m  V^2 }{2 \hbar^2}- \frac{ \lambda^2 }{2 m} .
\end{eqnarray}
Moreover, the normalization coefficient $A = \frac{\sqrt{m V}}{\hbar}$ can be given from
\begin{eqnarray}
\int^{+\infty}_{-\infty} |\psi(x)|^2 dx = 1.
\end{eqnarray}
At last, we obtain the corrected wave function expressed as
\begin{eqnarray}
\psi(x) = \frac{\sqrt{m V}}{\hbar}e^{-\frac{m V |x|}{{\hbar}^2}} e^{-\frac{i \lambda x}{\hbar}}.
\end{eqnarray}
\\
\\
\\
\section{One-dimension delta potential barrier}
In this section we will work out the effect of the the deformation on the delta potential barrier. we deform the momentum
$p\rightarrow p(1+\lambda/p)$ in the Hamiltonian and analyze the effects on a delta potential barrier. The delta potential barrier
is defined as
\begin{eqnarray}
V(x) =   V \delta(x),
\end{eqnarray}
and the corrected Hamiltonian of a delta potential barrier can be written as
\begin{eqnarray}
H = \frac{p^2}{2m} + V(x) = \frac{p^2}{2m} + \frac{\lambda p}{m} + V \delta(x).
\end{eqnarray}
From the corrected Hamiltonian it is easy to get the deformed Schr\"{o}dinger equation as follows
\begin{eqnarray}
\frac{\hbar^2}{2 m} \psi''(x) + i\hbar\frac{\lambda}{ m} \psi'(x) + \left[ - V \delta(x) +E \right]\psi(x) =0 . \label{deltapotentialbarrierequation1}
\end{eqnarray}
 Integrating the Eq.(\ref{deltapotentialbarrierequation1}) by $dx$ from$-\varepsilon$ to $\varepsilon$ and reducing  the positive $\varepsilon$ to infinitesimal, one can obtain the step equation of $\psi'(x)$
\begin{eqnarray}
 \psi'\left(0^+\right)-  \psi'\left(0^-\right) =   \frac{2 m V }{\hbar^2} \psi\left(0\right).
\end{eqnarray}
In the non-zero region of $x$, the Eq.(\ref{deltapotentialbarrierequation1}) becomes
\begin{eqnarray}
\frac{\hbar^2}{2 m} \psi''(x) + i\hbar\frac{\lambda}{ m} \psi'(x) +    E  \psi(x) =0.
\end{eqnarray}
By considering the left incoming wave it is noticed that  the solutions can be chosen in terms of
\begin{eqnarray}
 \psi(x) =\left\{
\begin{aligned}
 S e^{ \frac{i(\sqrt{ 2 m E + \lambda^2}- \lambda )x}{\hbar} }  \quad \quad  \quad \quad,   x>0 \\
 e^{ \frac{i(\sqrt{ 2 m E + \lambda^2}- \lambda )x}{\hbar} } + R e^{- \frac{i(\sqrt{ 2 m E + \lambda^2}+ \lambda )x}{\hbar} } ,   x<0  \end{aligned}
\right.
\end{eqnarray}
where $R e^{- \frac{i(\sqrt{ 2 m E + \lambda^2}+ \lambda )x}{\hbar} }$  stands for the reflected state  while $S e^{ \frac{i(\sqrt{ 2 m E + \lambda^2}- \lambda )x}{\hbar} }$ stands for  the transmitted state. According to the continuity of $\psi(x)$ and the step equation of $\psi'(x)$, the relations of $R$ and $S$ are
\begin{eqnarray}
&& 1 + R = S, \\
&& \frac{i(S-1)(\sqrt{2 m E +\lambda^2})}{\hbar} + \frac{i R (\sqrt{2 m E +\lambda^2})}{\hbar} = \frac{2 m V}{\hbar^2}S.
\end{eqnarray}
which gives $R$ and $S$
\begin{eqnarray}
S &=&   \frac{\hbar \sqrt{ 2 m E + \lambda^2}}{\hbar \sqrt{ 2 m E + \lambda^2} + i m V},   \label{transmittedstate } \\
R &=&  - \frac{ { i m V}}{\hbar \sqrt{ 2 m E + \lambda^2} + i m V}.
\end{eqnarray}
Then the wave functions  and  transmission coefficient $|S|^2$ and the reflectance coefficient $|R|^2$ are shown
\begin{eqnarray}
 \psi(x) =\left\{
\begin{aligned}
 \frac{\hbar \sqrt{ 2 m E + \lambda^2}}{\hbar \sqrt{ 2 m E + \lambda^2} + i m V} e^{ \frac{i(\sqrt{ 2 m E + \lambda^2}- \lambda )x}{\hbar} }    \quad \quad   \quad \quad,   x>0 \\
 e^{ \frac{i(\sqrt{ 2 m E + \lambda^2}- \lambda )x}{\hbar} } - \frac{ { i m V}}{\hbar \sqrt{ 2 m E + \lambda^2} + i m V}e^{- \frac{i(\sqrt{ 2 m E + \lambda^2}+ \lambda )x}{\hbar} } ,   x<0  \end{aligned}
\right.
\end{eqnarray}
\begin{eqnarray}
|S|^2 &=&   \frac{\hbar^2 \left( 2 m E + \lambda^2 \right) }{{\hbar^2 \left( 2 m E + \lambda^2 \right)+ m^2 V^2} }, \\
|R|^2 &=&   \frac{m^2 V^2 }{{\hbar^2 \left( 2 m E + \lambda^2 \right)+ m^2 V^2} }.
\end{eqnarray}
Consequently,  if we denote the usual transmitted state as  $S_0 e^{ \frac{i \sqrt{ 2 m E} } {\hbar} x}$, from Eq.(\ref{transmittedstate }) it is clear that
\begin{eqnarray}
S_0 =  \frac{\hbar \sqrt{ 2 m E }}{\hbar \sqrt{ 2 m E } + i m V},   \    \     \   |S_0|^2 =  \frac{\hbar^2 \left( 2 m E  \right) }{{\hbar^2 \left( 2 m E  \right)+ m^2 V^2} }.
\end{eqnarray}
Furthermore, the excess tunneling current  can be carried out as
\begin{eqnarray}
\frac{|S|^2 - |S_0|^2}{|S_0|^2} \approx \frac{\lambda^2  V^2}{2E (2  E \hbar^2 + m V^2)}.
\end{eqnarray}
\\
\\
\\
\section{One-dimension Coulomb potential}
We deform the Hamiltonian of the one-dimension Coulomb potential as
\begin{eqnarray}
H = \frac{p^2}{2m} -\frac{\kappa}{|x|} \rightarrow \frac{p^2}{2m} + \frac{\lambda p}{m} - \frac{\kappa}{|x|},
\end{eqnarray}
then  we obtain the deformed Schr\"{o}dinger equation
\begin{eqnarray}
\frac{\hbar^2}{2 m} \psi''(x) + i\hbar\frac{\lambda}{ m} \psi'(x) + \left( \frac{\kappa}{|x|} +E \right)\psi(x) =0, \label{Coulombequation1}
\end{eqnarray}
therefore we have gained the wave function  in terms of confluent hypergeometric function in the region  of $x>0$,
\begin{eqnarray}
\psi(x)= x e^{-\frac{i \lambda x}{\hbar}} e^{-\frac{\sqrt{-\lambda^2-2 E m}x}{\hbar}} F \left(1-\frac{\kappa m }{\hbar \sqrt{-\lambda^2 -2 E m}},2, \frac{2 x \sqrt{-\lambda^2 -2 E m} }{\hbar}\right).
\end{eqnarray}
With the boundary condition of bound states,
\begin{eqnarray}
|x| \rightarrow \infty, \psi(x)\rightarrow 0,
\end{eqnarray}
the confluent hypergeometric function must degenerate into a generalized Laguerre polynomial, in other words,
the  first parameter  of the confluent hypergeometric function $1-\frac{\kappa m }{\hbar \sqrt{-\lambda^2 -2 E m}}$  should be  a non-positive integer,
 \begin{eqnarray}
1-\frac{\kappa m }{\hbar \sqrt{-\lambda^2 -2 E m}} = 1-n,  n= 1, 2, 3, \cdots.
\end{eqnarray}
From this result it is clear that the energy level of the system can be described as
\begin{eqnarray}
E_n = - \frac{\kappa^2 m}{2 \hbar^2 n^2} -\frac{\lambda^2}{2 m},
\end{eqnarray}
and  we write the wave function
\begin{eqnarray}
\psi(x)= x e^{-\frac{i \lambda x}{\hbar}} e^{-\frac{\kappa m x}{n \hbar^2}} F \left(1-n ,2, \frac{2 \kappa m x  }{n \hbar^2}\right).   \label{GUPHydra1}
\end{eqnarray}
In the region of $x<0$ $\psi(x)$ satisfies
\begin{small}
\begin{eqnarray}
\psi(x)&=& x e^{-\frac{i \lambda x}{\hbar}} e^{-\frac{\sqrt{-\lambda^2-2 E m}x}{\hbar}} F \left(1+\frac{\kappa m }{\hbar \sqrt{-\lambda^2 -2 E m}},2, \frac{2 x \sqrt{-\lambda^2 -2 E m} }{\hbar}\right)     \nonumber \\
&=& x e^{-\frac{i \lambda x}{\hbar}} e^{\frac{\sqrt{-\lambda^2-2 E m}x}{\hbar}} F \left(1-\frac{\kappa m }{\hbar \sqrt{-\lambda^2 -2 E m}},2, -\frac{2 x \sqrt{-\lambda^2 -2 E m} }{\hbar}\right),
\end{eqnarray}
\end{small}
by repeating previous  procedures we get the same energy level  $E_n$ and the following wave function
\begin{eqnarray}
 x e^{-\frac{i \lambda x}{\hbar}} e^{\frac{\kappa m x}{n \hbar^2}} F \left(1-n ,2, -\frac{2 \kappa m x  }{n \hbar^2}\right). \label{GUPHydra2}
 \end{eqnarray}
 In the last,  because the two functions  Eq. (\ref{GUPHydra1}) and (\ref{GUPHydra2})  vanishes at the point $x=0$,   the  two possibilities of the whole wave function   must be taken  into account
  \begin{eqnarray}
 \psi_{A}(x) =\left\{
\begin{aligned}
 x e^{-\frac{i \lambda x}{\hbar}} e^{-\frac{\kappa m x}{n \hbar^2}} F \left(1-n ,2, \frac{2 \kappa m x  }{n \hbar^2}\right) ,   x>0 \\
- x e^{-\frac{i \lambda x}{\hbar}} e^{\frac{\kappa m x}{n \hbar^2}} F \left(1-n ,2, -\frac{2 \kappa m x  }{n \hbar^2}\right) ,   x<0  \end{aligned}
\right.
\label{doubledegeneratestate1}
\end{eqnarray}
and
 \begin{eqnarray}
 \psi_{B}(x) =\left\{
\begin{aligned}
 x e^{-\frac{i \lambda x}{\hbar}} e^{-\frac{\kappa m x}{n \hbar^2}} F \left(1-n ,2, \frac{2 \kappa m x  }{n \hbar^2}\right) ,   x>0 \\
x e^{-\frac{i \lambda x}{\hbar}} e^{\frac{\kappa m x}{n \hbar^2}} F \left(1-n ,2, -\frac{2 \kappa m x  }{n \hbar^2}\right) ,   x<0
 \end{aligned}
\right.
 \label{doubledegeneratestate2}
\end{eqnarray}
\\
\\
\section{One-dimension Stark effect }
In this section, we will analyze the effect of the deformation on a one-dimension hydrogen atom in an external electric field $\mathscr{E}$ along the coordinate axis in $x$ direction.
It is noted that the spectral lines of atoms will shift and split  when an external electric field imposed, which is called Stark effect.
We distinguishes first- and second-order Stark effects.
The Hamiltonian of the system can be corrected by the deformation as
\begin{eqnarray}
H = \frac{p^2}{2m} -\frac{\kappa}{|x|} + e \mathscr{E} x \rightarrow \frac{p^2}{2m} + \frac{\lambda p}{m} - \frac{\kappa}{|x|} + e \mathscr{E} x ,
\end{eqnarray}
whose Schr\"{o}dinger equation unfortunately has no analytical solution.  Now we use perturbation theory to calculate the energy corrections.  At the beginning, we choose
the  perturbation term  $H'$ of Hamiltonian and denote wave functions of the double degenerate states of $E_n^{(0)}$ ( Eq.(\ref{doubledegeneratestate1}) and (\ref{doubledegeneratestate2}) )  as
 \begin{eqnarray}
&& H_0 = \frac{p^2}{2m} + \frac{\lambda p}{m} - \frac{\kappa}{|x|}, \ \ \ \ \ \  \ \ \ \ \ \ \ \ \ \ \ \ \ \ \ \ \ \  H'=  e \mathscr{E} x ,  \\
&& \psi_{n1}^{(0)} \equiv  \psi_A(x) =\left\{
\begin{aligned}
 x e^{-\frac{i \lambda x}{\hbar}} e^{-\frac{\kappa m x}{n \hbar^2}} F \left(1-n ,2, \frac{2 \kappa m x  }{n \hbar^2}\right) ,   x>0 \\
- x e^{-\frac{i \lambda x}{\hbar}} e^{\frac{\kappa m x}{n \hbar^2}} F \left(1-n ,2, -\frac{2 \kappa m x  }{n \hbar^2}\right) ,   x<0  \end{aligned}
\right.   \\
 &&  \psi_{n2}^{(0)} \equiv \psi_B(x) =\left\{
\begin{aligned}
 x e^{-\frac{i \lambda x}{\hbar}} e^{-\frac{\kappa m x}{n \hbar^2}} F \left(1-n ,2, \frac{2 \kappa m x  }{n \hbar^2}\right) ,   x>0 \\
x e^{-\frac{i \lambda x}{\hbar}} e^{\frac{\kappa m x}{n \hbar^2}} F \left(1-n ,2, -\frac{2 \kappa m x  }{n \hbar^2}\right) ,   x<0
 \end{aligned}
\right.
\end{eqnarray}
We arrive at the results of first-order perturbation energy by solving the determinant
 \begin{eqnarray}
\left|
\begin{array}{cc}
 H'_{11}-E_n^{(1)} & H'_{12} \\
 H'_{21} & H'_{22}-E_n^{(1)} \\
\end{array}
\right| = 0,
\end{eqnarray}
where the elements of  this determinant $H'_{11},  H'_{22}$ are zero on account of  the parity of  the integrands
 \begin{eqnarray}
H'_{11} =  \langle  \psi_{n1}^{(0)} | H'  |  \psi_{n1}^{(0)} \rangle  =   e \mathscr{E}  \langle  \psi_{n1}^{(0)} | x  |  \psi_{n1}^{(0)} \rangle = 0.
\end{eqnarray}
 \begin{eqnarray}
H'_{22} =  \langle  \psi_{n2}^{(0)} | H'  |  \psi_{n2}^{(0)} \rangle  =   e \mathscr{E}  \langle  \psi_{n2}^{(0)} | x  |  \psi_{n2}^{(0)} \rangle = 0,
\end{eqnarray}
while one of the other two  elements of the determinant
\begin{eqnarray}
H'_{12} &=&  \langle  \psi_{n1}^{(0)} | H'  |  \psi_{n2}^{(0)} \rangle                \nonumber\\
& =&   e \mathscr{E}  \langle  \psi_{n1}^{(0)} | x |  \psi_{n2}^{(0)} \rangle     \nonumber\\
&=& 2 e \mathscr{E}   \int_0^{\infty} x^3 e^{-\frac{2 \kappa m x}{n \hbar^2}} \left[F \left(1-n, 2, \frac{2 \kappa m x}{n \hbar^2}\right)\right]^2 dx.
\end{eqnarray}
With the  definition of the generalized Laguerre polynomial $L_n^\mu (z)$ by confluent hypergeometric functions \cite{Laguerre}
 \begin{eqnarray}
L_n^\mu (z) \equiv \frac{\Gamma(\mu+1+n)}{n!  \Gamma (\mu+1)} F \left(-n, \mu+1, z\right)
\end{eqnarray}
we present $F \left(1-n, 2, \frac{2 \kappa m x}{n \hbar^2}\right)$  as
 \begin{eqnarray}
F \left(1-n, 2, \frac{2 \kappa m x}{n \hbar^2}\right) = \frac{1}{n} L_{n-1}^{1} \left(\frac{2 \kappa m x}{n \hbar^2} \right),
\end{eqnarray}
and using the equation \cite{Laguerre}
 \begin{eqnarray}
& &\int_0^{+\infty} z^\alpha e^{-z} L_{s}^{\beta}(z)  L_{t}^{\gamma}(z)  dz       \nonumber\\
&=& (-1)^{s+t} \Gamma(\alpha +1) \sum_k
\left(\begin{array}{lcl} {\alpha - \beta} \\   {s - k} \end{array}\right)
\left(\begin{array}{lcl} {\alpha - \beta} \\   {\gamma - k}  \end{array}\right)
 \left(\begin{array}{lcl} {\alpha + k } \\   \  \  \  k  \end{array}\right)
 \end{eqnarray}
 we  work out the result
\begin{eqnarray}
H'_{12} &=& \frac{2 e \mathscr{E} }{n^2} \int_0^{\infty} x^3 e^{-\frac{2 \kappa m x}{n \hbar^2}} \left[L_{n-1}^{1}\left(\frac{2 \kappa m x}{n \hbar^2} \right)\right]^2 dx
= \frac{3 e \mathscr{E} \hbar^8}{4 \kappa^4 m^4} n^5.
\end{eqnarray}
Similarly, another element of the determinant is
 \begin{eqnarray}
H'_{21} &=&  \langle  \psi_{n2}^{(0)} | H'  |  \psi_{n1}^{(0)} \rangle   = \frac{3 e \mathscr{E} \hbar^8}{4 \kappa^4 m^4} n^5 .
\end{eqnarray}
At last we have derived the first order corrections to the energy levels originating from the Stark effect are
 \begin{eqnarray}
E_{n1}^{(1)} =  \frac{3 e \mathscr{E} \hbar^8}{4 \kappa^4 m^4} n^5, \ \ \ \ \ \ \ E_{n2}^{(1)} = - \frac{3 e \mathscr{E} \hbar^8}{4 \kappa^4 m^4} n^5.
\end{eqnarray}
Meanwhile, the first-order approximation wave functions corresponding to the split energy levels are given,
 \begin{eqnarray}
\phi_{n1}^{(0)} =\frac{1}{\sqrt{2}} (\psi_{n1}^{(0)}  + \psi_{n2}^{(0)} )
 =\left\{
\begin{aligned}
\sqrt{2} x e^{-\frac{i \lambda x}{\hbar}} e^{-\frac{\kappa m x}{n \hbar^2}} F \left(1-n ,2, \frac{2 \kappa m x  }{n \hbar^2}\right) ,   x>0 \\
0 \quad \quad  \quad \quad \quad \quad  \quad \quad,   x<0
\end{aligned}
\right.
\end{eqnarray}
\begin{eqnarray}
\phi_{n2}^{(0)} =\frac{1}{\sqrt{2}} (\psi_{n1}^{(0)}  - \psi_{n2}^{(0)} )
 =\left\{
\begin{aligned}
0 \quad \quad  \quad \quad \quad \quad  \quad \quad, x>0 \\
-\sqrt{2} x e^{-\frac{i \lambda x}{\hbar}} e^{\frac{\kappa m x}{n \hbar^2}} F \left(1-n ,2, -\frac{2 \kappa m x  }{n \hbar^2}\right) ,   x<0
\end{aligned}
\right.
\end{eqnarray}

Thus, there are not degenerate energy states after the first order corrections, which shows that  the   nondegenerate perturbation theory can be  utilized
to calculate the second-order corrections to the energy levels
  \begin{eqnarray}
E_{n1}^{(2)} =  \langle  \phi_{n1}^{(0)} | H'  |  \phi_{n1}^{(0)} \rangle   = \frac{3 e \mathscr{E} \hbar^8}{4 \kappa^4 m^4} n^5
\end{eqnarray}
or
 \begin{eqnarray}
E_{n2}^{(2)} =   \langle  \phi_{n2}^{(0)} | H'  |  \phi_{n2}^{(0)} \rangle   =  -\frac{3 e \mathscr{E} \hbar^8}{4 \kappa^4 m^4} n^5.
\end{eqnarray}

Therefore, the two total energy levels of  a one-dimension hydrogen atom in an external electric field takes the form,
\begin{eqnarray}
 - \frac{\kappa^2 m}{2 \hbar^2 n^2} -\frac{\lambda^2}{2 m}+\frac{3 e \mathscr{E} \hbar^8}{2 \kappa^4 m^4} n^5,
 \end{eqnarray}
and
  \begin{eqnarray}
 - \frac{\kappa^2 m}{2 \hbar^2 n^2} -\frac{\lambda^2}{2 m}-\frac{3 e \mathscr{E} \hbar^8}{2 \kappa^4 m^4} n^5.
 \end{eqnarray}
\\
\\
\\
\section{Conclusion}
In this paper we first briefly review the most general deformation motivated by GUP in S. Masood's work. It is clear that the most general form of deformation has infinite inverse powers, and the Hamiltonian can bring many interesting results.

We have applied its simple limit to various one-dimension quantum models and analyzed the effects on them. We have found that the
corrected energy have the same form $-\lambda^2/{2 m}$, and  there is an extra phase factor emerging in every corrected wave function. In the framework of the GUP form, the symmetry of parity of the wave functions has been broken in consequence of the
linear term $\lambda p/m$. Especially, for one-dimension potential barrier the excess tunneling current being proportional to $\lambda^2$ can be checked by experiments.

Next, we want to apply the most general deformation of the momentum operator to other territories, for example black hole thermodynamics. It would be interesting to investigate the temperatures, the entropies and the heat capacities of a variety of black holes and it is to be expected that new physical features will emerge.

At last, we should try to use not only the simple limit but also more complex forms of the most general deformation GUP, which has made of great differences. In that situation, new physical results attendant on complicated calculations.

\section*{Acknowledgments}
Supported by Research Program of Qilu Institute of Technology (No.: QIT23NN036).

\end{document}